# High-Q dielectric Mie-resonant nanostructures
# (a mini-review)

Pavel Tonkaev[1] and Yuri Kivshar[1,2]

[1]ITMO University, 197101, St. Petersburg, Russia

[2]Nonlinear Physics Centre, Australian National University, Canberra ACT 2601, Australia

## Abstract

 Future technologies underpinning high-performance optical communications, ultrafast computations and compact biosensing will rely on densely packed reconfigurable optical circuitry based on nanophotonics. For many years, plasmonics was considered as the only available platform for nanoscale optics, but the recently emerged novel field of *Mie resonant metaphotonics* provides more practical alternatives for nanoscale optics by employing resonances in high-index dielectric nanoparticles and structures. In this mini-review we highlight some recent trends in the physics of dielectric Mie-resonant nanostructures with high quality factor (Q factor) for efficient spatial and temporal control of light by employing multipolar resonances and the bound states in the continuum. We discuss a few applications of these concepts to nonlinear optics, nanolasers, subwavelength waveguiding, and sensing.

**Introduction**

Nanophotonics is considered often as a special branch of optics that studies the behaviour of light on the nanometer scales including interaction of subwavelength objects with light. For many applications and for creating compact optical circuits and networks, it is highly desirable to miniaturize photonic components, and traditionally nanophotonics was based on metallic components which transport and focus light via surface plasmon polaritons which allow to overcome the diffraction limit [1]. However, plasmonic components are known to suffer from strong dissipative losses and heating. Recently, we have observed the emergence of a new field of *all-dielectric resonant metaphotonics* [2] (also called "*Mie-tronics*" [3]) aiming at the manipulation of strong optically-induced electric and magnetic Mie-type resonances in dielectric nanostructures with high refractive index. Unique advantages of such dielectric resonant nanostructures over their metallic counterparts are low dissipative losses combined with the strong enhancement of both electric and magnetic fields, thus providing competitive alternatives for plasmonics including optical nanoantennas, biosensors, and metasurfaces.

High-index dielectric nanoantennas supporting multipolar Mie resonances represents a novel type of building blocks of metamaterials for generating, manipulating, and modulating light. By combing both electric and magnetic multipolar modes, one can not only modify far-field radiation patterns but also localized the electromagnetic energy in open resonators by employing the physics of bound states in the continuum (BICs) to achieve destructive interference of two (or more) leaky modes [4]. Optical Mie resonances in nanoantennas can be characterized the average lifetime of trapped light being quantified by the value of the quality factor (Q factor). Lower-order dipolar Mie modes are known to have relatively low Q factors, of the order of ten. Changing the resonator parameters or combining the resonators into a planar geometry of metasurfaces allow achieving much higher values of the Q factor.

This mini-review aims to highlight some recent advances in the field of *all-dielectric Mie-resonant metaphotonics* driven by the development of high-Q dielectric structures for nonlinear nanophotonics, nanoscale lasing, and efficient sensing applications.

Figure 1(a-e) show the examples of the SEM images (adopted from Refs. [6-10]) of several types of dielectric structures. In particular, subwavelength optical antennas made of high-index dielectric materials can have various shapes such cubes and cylinders [see Figs. 1(a, b)], and they support multipolar electric and magnetic Mie resonances which can be hybridized to realize *quasi-BIC resonances* [5] with high values of the Q factor. Below, we discuss how such Mie-resonant nanoantennas can be employed for second-harmonic generation and also for creating smallest non-plasmonic lasers operating at room temperatures. Figures 1(c, d) present arrays of dielectric nanoparticles for slow-light waveguides. Finally, combining subwavelength resonators into metasurfaces [see Fig. 1(e)] allows to employ collective resonances related to the BIC physics, and thus create a novel platform for nonlinear optics.

Importantly, various degrees of freedom and geometries can be employed to engineer the Q factor of such resonant dielectric structures, and also create slow-light waveguides based on the Mie resonances. Achieving high-Q resonances in metasurfaces provides novel strategies for creating compact optical devices for various applications in biosensing.

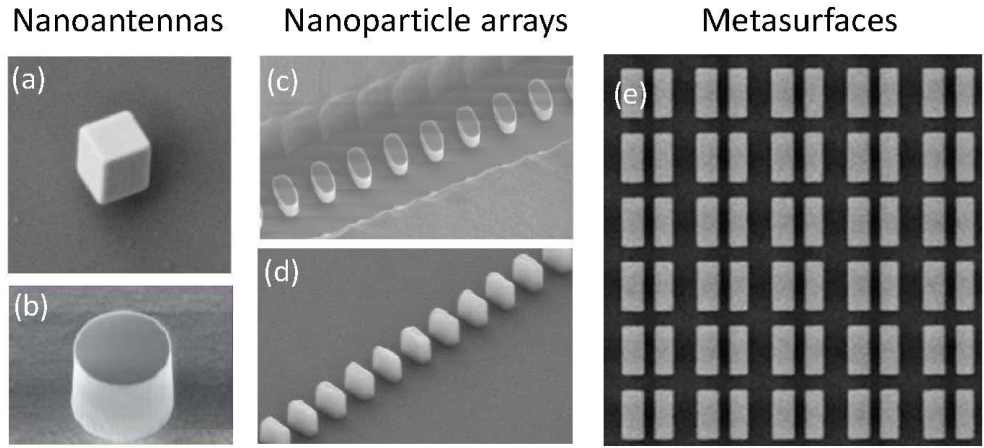

| Nanoantennas | Nanoparticle arrays | Metasurfaces |

**Fig. 1**. Major types of subwavelength dielectric structures shown as SEM images of the fabricated samples. (a, b) Cube and cylinder optical antennas employed for lasing and harmonic generation; (c, d) arrays of dielectric nanoparticles for slow-light waveguides; and (e) dielectric metasurfaces supporting BIC resonances. The images are adopted from Refs. [6-10].

## Nonlinear nanoantennas

Subwavelength dielectric nanoantennas made of high index materials emerged recently as a new platform for nanophotonics [2, 7, 11]. However, the enhancement of near-field effects for individual subwavelength resonators is strongly limited by low Q factors of the fundamental resonances governing the optical response. Recently, a novel approach [7] suggested how to achieve high-Q resonances in individual subwavelength resonators in the regime of *supercavity mode*, by employing the physics of nonradiative states – optical *bound states in the continuum* (BICs). Supercavity modes attracted a lot of attention, but they have been observed experimentally only this year. Koshelev et al [7] reported on the first experimental observation of the supercavity modes in individual subwavelength dielectric resonators, and they also demonstrated the record-high efficiency of the second-harmonic generation. That study revealed that supercavity modes are formed due to strong coupling of two leaky modes excited simultaneously in a dielectric resonator, which interfere destructively resulting in strong suppression of radiative losses. Those observations confirmed that the supercavity modes are governed by the physics of bound states in the continuum. This work opens novel opportunities for subwavelength dielectric metaphotonics and nonlinear nanophotonics.

More specifically, Koshelev et al [7] considered cylindrical resonators with height 635 nm made of AlGaAs ($\varepsilon$=11) placed on a silica substrate with an additional highly doped 300 nm ITO layer. This layer provides an additional enhancement of the Q factor due to the interaction between the resonator and substrate. To engineer the supercavity mode, they vary the resonator diameter between 890 nm and 980 nm to induce strong coupling between a pair of leaky Mie-resonant modes [7]. For an efficient excitation, the authors employ a tightly focused azimuthally polarized vector beam with the wavelength varying from 1500 nm to 1700 nm. The measured Q factor is about 190, which is more than one order of magnitude higher than for conventional magnetic dipolar Mie resonance. This high value of the Q factor stipulates the enhance of light-matter interaction observed through nonlinear effects. Figure 2(a) shows the measured second-harmonic intensity vs. particle diameter and incident wavelength excited with a structured pump. The inset shows the far-field profiles of the emitted radiation originating from the excited quasi-BIC mode.

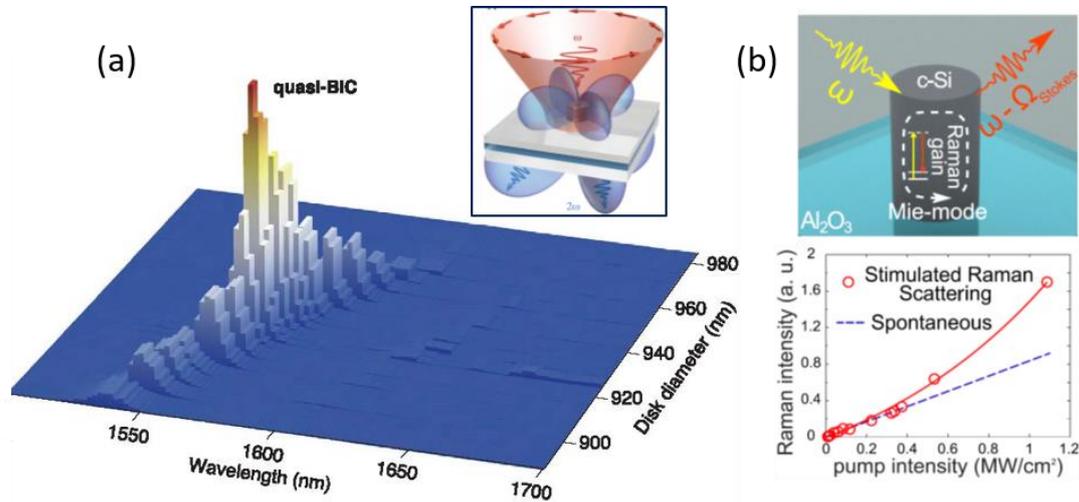

**Fig. 2**. Examples of nonlinear effects enhanced by Mie resonances. (a) Generation of the second harmonic empowered by the quasi-BIC resonance. (b) Stimulated Raman scattering from c-Si nanopillars driven by a low-order Mie resonance. Adopted from Refs. [7,11].

Raman scattering has been demonstrated as an effective tool to study optical modes. A significant increase of the Raman scattering signal corresponds to the effective energy coupling with modes of the nanoparticle. Spontaneous Raman scattering enhanced by the electric and magnetic Mie resonances of subwavelength particles [12] can be used for sensing [13] and nanothermometry [14]. However, spontaneous Raman scattering remains a relatively weak effect as compared with photoluminescence observed in resonant nanostructures [15]. Zograf et al. [11] observed experimentally *stimulated Raman scattering* from isolated subwavelength c-Si nanoparticles empowered by multipolar Mie resonances. They studied the scattering of a c-Si nanopillar with height of 600 nm and diameters ranging from 200 nm to 1000 nm fabricated by electron-beam lithography on $Al_2O_3$ substrate [see Fig. 2(b)].

First, the authors selected the subwavelength nanoparticle with the strongest Raman scattering signal. Confocal Raman scattering measurements from individual c-Si nanoparticles at a pump wavelength of 633 nm were obtained to achieve this goal. They found that at low intensity the maximum Raman scattering signal corresponds to the nanoparticle diameter of 475 nm driven mostly by the magnetic dipole and magnetic quadrupolar modes at 654 nm and 633 nm wavelengths. Typical experimental dependencies of the Raman scattering signal on the pump intensity are shown on Figure 2(b). At low intensities, the dependence is linear and spontaneous Raman scattering dominates. However, at higher intensities (higher than 0.3 MW/cm$^2$), a nonlinear growth of the Raman scattering signal was observed for the nanodisk with a diameter of 475 nm, which is not detected for other nanodisks with diameters from 250 nm to 800 nm. This sharp difference characterizes the stimulated Raman scattering process.

In the regime of the stimulated Raman scattering, a stronger pump transfers a part of its energy to a weaker Stokes-shifted Raman signal by interacting with a nonlinear medium. As a result, the pump can amplify a weak Stokes beam, attributing to the Raman amplification. When the pump photons become trapped by a resonator mode, the Raman scattering gets enhanced, and it generates a signal that grows exponentially with the input intensity.

**Non-plasmonic nanolasers**

Creation of nanoscale source of coherent radiation is an important part in the development of future optical computer systems. Until now, fabrication of semiconductor lasers has been limited to a few microns. Further size reduction becomes complicated due to radiation losses. The use of plasmonic materials can reduce radiation losses and laser size, but also increase non-radiative losses [16]. Recently, resonant dielectric nanoantennas have been shown as a platform for nanophotonic applications. Dielectric particles have low non-radiative losses that make them more attractive for lasing. The use of dielectric nanoantennas is aimed at enhancing the Raman signal [12] and photoluminescence of the nanoantenna material itself [17].

Halide perovskites are one of the best materials for lasing. This material has high enough refractive index that make possible creation of a compact design [15,17]. In addition, lead halide perovskites have low concentration of defects and high photoluminescence quantum yield [18]. Moreover, simple chemical fabrication methods allow creating optical resonant microstructures which generate stimulated emission in the optical frequency range [19,20]. However, the creation of smaller resonators is challenging due to the fact that smaller particles have larger surface-to-volume ratios, which increases the effect of surface recombination and roughness and result in higher losses. Recent studies have overcome this limitation creating the 310 nm nanolaser based on lead halide perovskite and operating at room temperature with 0.58 size/wavelength ratio [6]. Nanoparticle $CsPbBr_3$ was synthesized chemically on $Al_2O_3$ substrate [see Fig. 3(a)]. Multipole decomposition of the lasing mode demonstrate the dominant contribution of the third-order magnetic dipolar Mie mode [6].

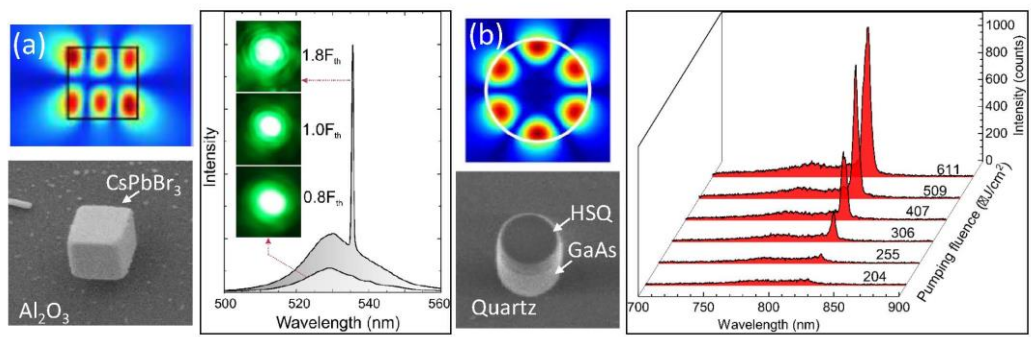

**Fig. 3**. Non-plasmonic nanolasers governed by low-order Mie-resonant modes in (a) cubic perovskite nanoparticle and (b) GaAs nanocylinder. Adopted from Refs. [6,21].

Creation of smaller resonators while keeping the same operating wavelength implies necessarily the use of lower-order modes. Those present higher radiative losses lead to a drop in the Q-factor. Mylnikov et al. [21] proposed a cylindrical nanoscale resonator supporting quasi-BIC mode [see Fig 3(b)]. The use of the regime of a supercavity mode reduces radiative losses enabling gain and loss compensation in nanoparticles of a smaller size. As a result, the authors demonstrated a GaAs nanolaser as small as 500 nm in diameter and only 330 nm in height on the quartz substrate with the lasing wavelength 825 nm, with the size-to-wavelength ratio 0.6 at cryogenic temperatures [21].

### Slow-light waveguides

Recent progress in nanoscale fabrication technologies provides more opportunities for a design of subwavelength metamaterial-inspired structures with engineered optical properties which are expected to enhance the performance of the next-generation photonic devices. Subwavelength guiding of light has attracted great attention because it provides unique opportunities for miniaturization of the optical interconnect technology. Recently, diverse implementation of subwavelength-engineered structures in integrated optics has been discussed for a design of integrated photonic platforms [22]. The major goals of those studies are to engineer the mode dispersion and waveguide anisotropy and, in particular, achieve slow-light propagation. The next step would be to employ resonances in such slow-light waveguides based on both electric and magnetic Mie-multipolar modes [23].

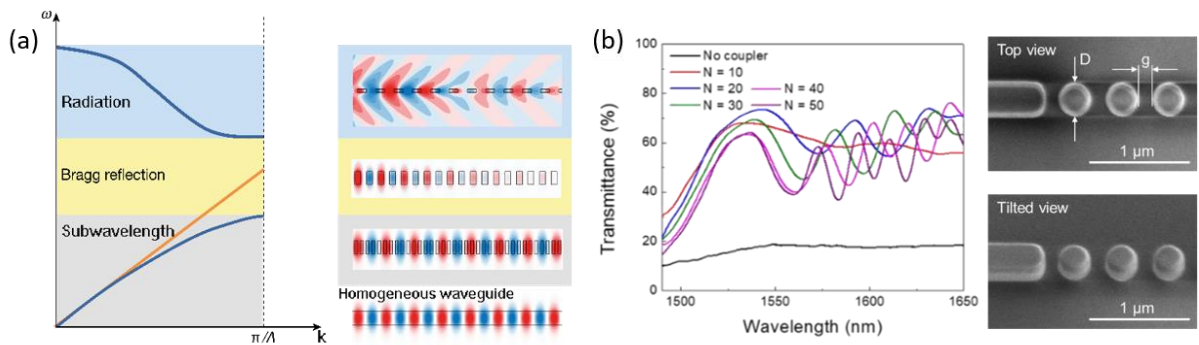

**Fig. 4.** Subwavelength waveguiding. (a) Schematic dispersion (left) and corresponding electric field profiles (right) of a periodic slab waveguide for three regimes: radiation, Bragg scattering, and subwavelength-guided wave propagation. (b) Transmittance and SEM images of waveguiding arrays of Mie-resonant silicon nanoparticles. Adopted from Refs. [22,23].

In general, light propagation through a periodic slab waveguide is governed by the dispersion shown in Fig. 4(a) (left) with radiation, Bragg scattering, and subwavelength regimes [22]. In the radiation regime, the periodic waveguide represents a diffraction grating radiating from the waveguide into free space above and below, as seen in Fig. 4a (right). For shorter periods, the waveguide supports Floquet–Bloch modes characterized by an electric field modulated with the same periodicity as the waveguide. However, when the waveguide periodicity becomes substantially smaller than the wavelength, the waveguide can be effectively approximated as a uniaxial crystal, thus enabling engineering of the refractive index, as discussed in Ref. [22]. As a result, the waveguide is optically equivalent to a homogeneous waveguide with an effective core index determined by the filling factor (the red line shows the effective dispersion). The ability to control both dispersion and anisotropy of waveguides provides a powerful design tool to engineer the wavevectors of the propagating modes [22].

However, the situation changes dramatically when the properties of individual resonators are employed. Earlier theoretical studies [24] revealed that the electromagnetic energy can be guided efficiently along a chain of nanoparticles and even around a corner, when the field is localized on the size less than half of the guided wavelength, thus providing an alternative to plasmonic waveguides. This concept has been verified at microwaves [25] using an analogue of optical nanoantennas realized at microwaves [26]. These results suggested that a chain of high-index nanoparticles can guide light for the distances exceeding substantially the guiding distances achievable with metallic or metal-dielectric plasmonic waveguides.

High-index dielectric nanoparticles with Mie resonances open new opportunities to control light at the nanoscale, and they can support slow-light waveguiding in a chain of coupled Mie-resonant silicon nanoparticles at telecommunication wavelengths. Recent studies demonstrated experimentally the opportunities and advantages of these novel types of subwavelength waveguides and their applications [27,28].

Figure 4(b)(left) shows transmittance of the specific optical device employing a chain of nanoparticles, shown in Fig. 4(b)(right) with the diameters D =340 nm and array spacing g = 150 nm, as a hybrid tapered coupler [23]. The black curve is the device transmittance without the nanoparticle coupler, while coloured curves are for the devices with couplers of various tapered lengths defined by the number of nanoparticles N in the taper. One observes that the overall transmittance through the device increases from 20% to ~70% due to the nanoparticle coupler. Also, such Mie-resonant arrays may show even lower propagation losses compared to that of stripe waveguides with 400 nm width, while having smaller footprints, and they reduce the group velocities of guided modes down to 0.03 of the speed of light.

**High-Q dielectric metasurfaces**

Surface-enhanced spectroscopy techniques are usually employed for extracting chemically specific information that is intimately linked to molecular structure and conformation. Traditionally, surface-enhanced spectroscopy is based on metallic nanoantenna supporting plasmon resonances, but it faces significant limitations. Dielectric metasurfaces emerged as an alternative approach for a control of both electric and magnetic fields with low material losses, and they can provide CMOS compatibility for large-scale manufacturing. Recently, high-Q metasurfaces have been suggested for mid-IR spectroscopy to encode spatially molecular absorption signatures into chemically specific barcodes [29]. High-Q resonances are created by using symmetry-broken resonator geometries supporting bound states in the continuum [30]. By creating pixelated metasurface structure, one may assign different resonances to specific pixels, mapping both spectral and spatial information. The strength of the molecular absorption signatures correlates with reflectance signal variations for different pixels, allowing a read-out in an imaging-based setup. Such a molecular imaging can be performed by using broadband light sources and detectors, thus enabling spectrometer-less operation in a miniaturized platform for on-site applications. Multi-component samples containing biomolecules, environmental pollutants, and polymers can be analysed by comparing the barcode of the unknown mixture with a library of reference barcodes.

Angle multiplexing is a powerful concept that allows encoding of different values of optical parameters. Figure 5(a) shows an angle-multiplexed metasurface [29] designed to resonantly reflect in a narrow spectral range around a frequency at each incidence angle when illuminated with a broadband source. This optical response is provided by a high-Q dielectric metasurface consisting of anisotropic arrays of germanium resonators on a calcium fluoride (CaF$_2$) substrate, which interact collectively to generate high-Q resonances in reflection.

Another useful direction to employ high-Q dielectric metasurfaces is to combine them with atomically thin monolayers of transition metal dichalcogenides (TMDCs). Intrinsic nonlinearity of TMDC monolayers is weak, and thus this limits their applications in nonlinear optics. However, the effective nonlinear susceptibility of TMDCs can be enhanced substantially by integrating them with dielectric metasurfaces supporting bound states in the continuum. Recently, Bernhardt et al [31] demonstrated that a WS$_2$ monolayer combined with a silicon metasurface hosting BICs exhibits second-harmonic generation enhanced by more than 3 orders of magnitude relative to a WS$_2$ monolayer placed on top of a flat silicon film of the same thickness. These results suggest a pathway to employ high-index dielectric metasurfaces as hybrid structures for the enhancement of TMDC nonlinearities with applications in nonlinear microscopy and optoelectronics.

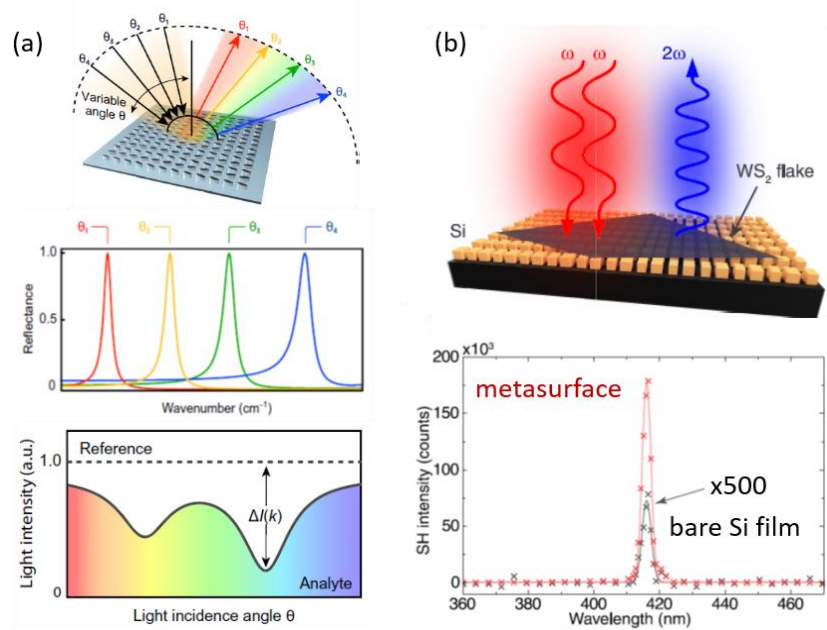

**Fig. 5**. Applications of BIC-empowered high-Q dielectric metasurfaces. (a) Concept of the angle-multiplexed sensing based on a dielectric metasurface. (b) Second-harmonic generation from a WS$_2$ monolayer placed on top of a Si metasurface, and the intensity spectra of the generated second-harmonic signal. Adopted from Refs. [29,31].

Figure 5(b) shows schematically a hybrid photonic structure (top) for enhanced nonlinear effects composed of a WS$_2$ monolayer placed on top of a silicon metasurface created by a square array of nanobar pairs. The image below shows the measured second-harmonic intensity spectra from WS$_2$ monolayers placed on top of the optimized metasurface (red) and on top of the reference bulk Si film (grey, magnified 500×) with a pump wavelength of 832 nm.

## Concluding remarks

Low-loss dielectric subwavelength structures and metasurfaces supporting high-Q resonances are ideally suited for applications in nanophotonics, including high-harmonic generation, biosensing, and quantum effects. Mie-resonant nanoantennas and BIC-resonant metasurfaces can enhance nonlinear response of hybrid materials when such materials are placed within the near-field of the Mie resonances, being useful for enhanced functionalities of novel two-dimensional materials and their hybrid structures. More importantly, active Mie-resonant nanoantennas can be employed as the smallest light sources with the footprints several times smaller than the wavelength of the light with a perspective of creating dense photonic integration for efficient on-chip metadevices.

Novel technologies based on smart engineering of multipolar Mie resonances and bound states in the continuum may enhance substantially light-matter interaction creating resonant linewidths for practical optical devices. Combining the advantages of flat optics with a platform of dielectric metasurfaces could enable a new strategy for achieving tunable control over optical wavefronts with an electromagnetic field. As one of the breakthrough applications, we would like to mention a novel type of optical biosensors and chiral sensors employing high-Q resonances in transmissive and reflective dielectric structures, and thus extending the radiation channels available for sensing increasing both device sensitivity and multiplexing abilities.

Modern integrated photonics requires the developments in device design, materials synthesis, nanofabrication, and characterization, and the combination of these efforts will underpin new discoveries and applications ranging from flat lenses to quantum computation and storage.

## Acknowledgements

The authors acknowledge a valuable collaboration and discussions with their colleagues and co-authors, especially K. Koshelev, S. Makarov and M. Rybin. This work was supported by the Russian Science Foundation (project № 20-73-10183), the Australian Research Council (grant DP200101168), and the Strategic Fund of the Australian National University.